\documentstyle[12pt]{article}
\def\bi{\bibitem}
\newcommand{\be}{\begin{equation}}
\newcommand{\ee}{\end{equation}}
\newcommand{\beq}{\begin{eqnarray}}
\newcommand{\eeq}{\end{eqnarray}}
\newcommand{\bear}{\begin{array}}
\newcommand{\ear}{\end{array}}

\begin{document}
\title{ Schwinger's Result On Particle Production From Complex Paths WKB 
Approximation}    
\author{
S.Biswas$^{*a),b)}$, A.Shaw $^{**a)}$ and B.Modak$^{a)}$ \\ 
a) Dept. of Physics, University of Kalyani, West Bengal,\\
India, Pin.-741235 \\
b) IUCAA, Post bag 4, Ganeshkhind, Pune 411 007, India \\
$*$ email: sbiswas@klyuniv.ernet.in\\
$**$ email:amita@klyuniv.ernet.in}
\date{}
\maketitle
\begin{abstract}
This paper presents the derivation of Schwinger's gauge invariant result of 
$Im\;{\cal {L}}_{eff}$ upto one loop approximation, for particle production in 
an uniform electric field through the method of complex trajectory WKB 
approximation (CWKB). The CWKB proposed by one of the author's \cite{bis:ijtp} 
looks upon particle production as due to the motion of a particle in complex 
space-time plane, thereby requiring tunneling paths both in space and time. 
Recently \cite{srini:iucaa,srini1:iucaa} 
there have been some efforts to calculate the reflection and transmission 
co-efficients for particle production in uniform electric field that differ 
from our expressions for the same. In this paper we 
clarify the confusion in this regard and establish the correctness of CWKB.
\end{abstract}
\section{\bf{Introduction}}
The complex trajectory WKB approximation has been an effective tool to understand
particle production. It works in Minkowski and as well as in curved spacetime;
both in space and time dependent gauge. Though originally proposed and 
developed by one of the authors \cite{bis:ijtp}, the method is now being persued by some 
workers \cite{srini:iucaa,srini1:iucaa,brout:pr}, technically being the same in approach with CWKB, having a 
slight differences with respect to interpretation and results. To show the 
exactness of CWKB with the standard results, we consider the particle production 
in constant electric field, where gauge invariant result due to Schwinger 
\cite{sch:pr} is available for comparison.
\par
Schwinger \cite{sch:pr} demonstrated that the probability for a vacuum to remain a 
vacuum in a volume $L^3=V$ over time $T$ is determined by the quantity 
$\exp{(-2Im\,{{\cal L}_{eff}})}V T$, i.e., $2Im\,{{\cal L}_{eff}}$ is the rate of 
decay of the vacuum per unit 4-volume. In one loop approximation, the 
Schwinger's result for a constant and uniform electrical field is 
\be
2Im\,{{\cal L}_{eff}^{(1)}}
=(2s+1){{(qE_0)^2}\over {(2\pi)^3}}
\sum_{n=1}^{\infty}{{(\pm)^{n+1}}\over n^2}e^{{-\pi n m^2}\over {qE_0}}.
\ee
It is a gauge invariant result. Here $m$ is the mass of the particle produced
with charge $q$ 
(for fermions ($s=1/2$) the upper sign, and for bosons ($s=0$) the lower sign is 
to be taken in Eq. (1)), and $E_0$ is the electric field. The 
expression (1) is related to pair production through Nikishov's 
\cite{nar:leb} representation : 
\be
2Im\,{{\cal L}_{eff}^{(1)}}L^3T=\mp\sum_{s}\int{{{d^3k}\over {(2\pi)^3}}L^3
\ln{(1\pm{\bar{n}}_p)}},
\ee
\be
{\bar{n}}_p=\exp{(-\pi{{m^2+k_{\bot}^{2}}\over {qE_0}})},
\ee
where ${\bar{n}}_p$ is the average number of pairs formed by the field in the state 
with a given momenta and projection of spin $k={\vec{k}},s$.
\par
The above technique of obtaining the $Im\,{\cal L}_{eff}$ can be generalized to 
arbitrary electromagnetic and gravitational backgrounds but is an uphill task 
since it is not at all convenient to regularize the effective Lagrangian in 
all such cases. Due to these reasons many workers in this field turn towards the 
Bogolubov transformation technique based on the method of normal mode analysis 
to understand the phenomenon of particle production. Here the problem arises 
with respect to interpretation when one tries to obtain the gauge invariant 
results, mentioned in (1) and (2). In time dependent gauge the 
instability of the vacuum is quite understandable. The vacuum $\vert 0_->$ at
$t\rightarrow -\infty$ 
is not the same as the vacuum  
$\vert 0_+>$ at 
$t\rightarrow +\infty$. 
The vacuum decay 
$\vert 0_->\rightarrow \vert 0_+>$ allows one to see particle production in 
which the Bogolubov coefficient $\beta_\lambda$ ($\lambda$ is a parameter that 
specifies the state of produced particles) is not zero and is interpreted as  
\be
<0_-\vert N(\lambda)\vert 0_+>=\sum_\lambda\vert\beta_\lambda\vert^2,
\ee
where $N(\lambda)$ is the number of `in' $u_i$-mode particles found in the out vacuum
$\vert 0_+\,>$, where $u_{i}{(\lambda)}$ modes construct the in-vacuum 
$\vert 0_->$ and ${\bar{u}}_{i}(\lambda)$
modes construct the out-vacuum $\vert 0_+>$. When one considers the same 
problem in space dependent gauge; the vacuua at $x\rightarrow \mp \infty$ 
remain the same and there is no vacuum instability. Consequently one expects no 
particle production in a space dependent gauge. However through mode analysis 
one still recovers a non-zero $\vert\beta_\lambda\vert^2$ even in space 
dependent gauge, implying particle production. A tunneling interpretation is then 
invoked to interpret particle production in space dependent gauge because of 
non-zero $\vert\beta_\lambda\vert^2$. In this interpretation one switches to the 
interpretation of `tunneling through the barrier' \cite{brout:pr} to calculate the 
reflection and transmission amplitude $R$ and $T$ and to obtain 
$\vert\beta_\lambda\vert^2$ through a prescription, identifying 
$\vert\alpha_\lambda\vert^2\pm\vert\beta_\lambda\vert^2 =1$ as being equivalent to 
$\vert T\vert^2\mp\vert R\vert^2=1$ ($\pm$ refer to spinor/scalar fields). The
prescription that relates $R$ and $T$ with $\alpha_\lambda$ and $\beta_\lambda$ 
is found to work differently in space and time dependent gauges. Recently a 
complex path analysis based on a technique of Landau \cite{lan:quant} is being used 
\cite{srini:iucaa,srini1:iucaa} to obtain $\beta_\lambda$ from the expressions of $R$ and $T$. In CWKB,
we used the technique of ref. \cite{knoll:ann} using the idea of Cornwall and Tiktopoulos 
\cite{cor:prd}. 
As pointed by us earlier, this has now led many to realize the usefulness of 
complex path analysis to 
understand particle production. In our view, the analytic continuation inherent 
in Bogolubov's technique is exploited in ref. \cite{brout:pr} with an advantage towards 
the point of interpretation. However these works do not clarify the source of 
instability, how and where the particle production occurs. In these works
\cite{srini:iucaa,srini1:iucaa,brout:pr} one also requires an advance knowledge 
of the gauge invariant result to 
settle the relation between the pairs $(R,\,T)$ and 
$(\alpha_\lambda,\beta_\lambda)$. The knowledge of mode solutions for the 
analytic continuation $x\rightarrow -\infty$ to $x\rightarrow +\infty$ 
(or, $t\rightarrow -\infty$ to $t\rightarrow +\infty$) is a basic requirement in 
these works. In principle one should obtain $N(\lambda)$, without connecting it 
to $\beta_\lambda$, from the results of $R$ and $T$ themselves. There is another 
point of significant argument that has motivated the present work. In CWKB, the 
results of $R$ and $T$ do not coincide with ref. 
\cite{srini:iucaa,srini1:iucaa,brout:pr} and hence needs discussion. In ref. 
\cite{srini:iucaa,srini1:iucaa}, 
attempt has been made to obtain the results (1) and 
(2), but being partially successful. In the present paper we use CWKB to obtain the 
reflection amplitude $R_{\sc{CWKB}}$ and transmission amplitude $T_{\sc{CWKB}}$ 
and use a first 
principle derivation of $N(\lambda)$ without reverting to the calculation of 
Bogolubov co-efficient and obtain the exact results, term by term, of (1) and 
(2) along with the prefactor in the expression (1). In CWKB, we do not require to 
know the mode solutions at $t\rightarrow \pm\infty$ (or $x\rightarrow \pm\infty$).
It is observed that $R_{\sc{CWKB}}$ is related to the full $S$-matrix elements
whereas $R_{\sc{C}}$, in the works of ref. \cite{brout:pr} is related to the connected 
$S$-matrix element.
\par
In CWKB, the particle production is considered as a `reflection' process either 
in time (in time dependent gauge) or in space (in space dependent gauge). This 
is the well known Klein paradox like situation, found in all standard textbooks 
where the pair creation occurs in a barrier region due to reflection causing 
Zitterbewegung within wavepacket in the barrier region. Only when the potential 
is capable of transferring energy at least $2mc^2$ to the wave packet, i.e., the 
minimum energy required to conserve energy in making a pair, will then real 
``permanent'' particles be created with the packet moving past the potential. 
The Zitterbewegung region is specified by two points within which the mixing of 
positive and negative frequency parts takes place and the interference term 
varies in time with a frequency at least $2mc^2$. The `two points' are 
the turning points from where the particle turns back. The turning points for 
a given problem is fixed by the one dimensional Schroedinger-like equation 
either in space or in time depending on the gauge chosen.
\par
The present work is motivated to deal with particle production through the 
technique of complex trajectory WKB (CWKB) approximation, proposed by us in 
various earlier works \cite{bis:pram,bis1:pram,bis:cqg,bis1:cqg} 
(mainly restricted to particle production in expanding 
spacetime). The organization of the paper is as follows. In section II we 
discuss the particle production as a process of reflection in time. It is a 
Klein-Paradox-like situation not in space but in time. We choose an idealised 
example to elucidate our approach. In section III we show a method to calculate
the number of produced particles in terms of reflection coefficient and obtain 
the Schwinger's result. In section IV we obtain the relation connecting the 
Bogolubov coefficients and the reflection coefficients. The concluding section 
synopsizes the results obtained in the present work and clarify some 
misconception on the CWKB results.
\section{\bf{Basic Principles of CWKB}}
The method of CWKB in the context of cosmology was proposed by one of the 
authors in \cite{bis:ijtp} and subsequently developed in 
\cite{bis:pram,bis1:pram,bis:cqg,bis1:cqg,sar:pram,sar:ijmpa} to treat particle 
production in Robertson-Walker spacetime and deSitter spacetime. In Feynman 
spacetime diagram a pair production is looked upon as 
$Vac\; \rightarrow e^++e^-$, in which the vacuum supplies energy to 
create the pair $2mc^2$ such that the positron $e^+$ of energy $E_{e^+}$ and the 
electron of energy $E_{e^-}$ both move forward in time. Using 
Feynman-Stuckleberg prescription, the electron moving forward in time can be 
considered as positron of energy $-E_{e^-}$ moving backward in time. The net 
result is that the pair production is looked upon as a process of reflection, 
not in space but in time from vacuum \cite{cor:prd}. The change 
$-E_{e^-}$ to $+E_{e^-}$ requires that the system showing pair production 
through process of reflection must have turning points in time and as the system 
evolves from $t\rightarrow -\infty$ to $t\rightarrow +\infty$, there is rotation 
of currents showing antiparticle to particle rotation. Obviously the turning 
points are complex in time, reflecting the situation of `over the barrier 
reflection'.
\par 
To get an idea of the above mechanism consider an one dimensional Schroedinger 
like equation in time
\be
{{d^2\psi}\over {dt^2}}+\omega^2(t)\psi=0.
\ee
Eq.(5) has turning points given by 
\be
\omega^2(t_{1,2})=0
\ee
where the turning points $t_{1,2}$ are complex. Balian and Bloch \cite{bal:ann} 
formulated an approach to quantum mechanics starting from classical trajectories 
with complex coordinates. In the approach, within the framework of WKB 
approximation, wave optics is generalized to complex trajectories to build up 
the wave. Such a method accounts for contribution of $\exp{(-{c\over\hbar})}$ to 
the usual WKB wave and is found to reproduce quantitively all quantum mechanical 
effects even in cases where the scattering potential varies rapidly over a 
distance of wavelength or less. We applied this technique following \cite{knoll:ann}, 
to obtain an expression for $\psi (t)$ at $t\rightarrow\infty$ as  
\be
\psi (t)\matrix{\longrightarrow \cr t\rightarrow\infty}
{1\over {(\omega (t))^{1\over 2}}}
\left[e^{iS(t,t_0)}-iRe^{-iS(t,t_0)}\right]
\ee
where
\be
S(t,t_0)=\int_{t_0}^{t} \omega (t) dt
\ee
and
\be
R={{\exp{\left[ 2iS(t_1,t_0)\right]}}\over 
{1+\exp{\left[ 2iS(t_1,t_2)\right]}}}.
\ee
The detailed derivation of Eq. (7) will be found in \cite{knoll:ann} and also in our 
previous works \cite{bis:cqg,bis1:cqg}. The interpretation of Eq. (7) is as follows. A wave 
starts at $t_0>t$ and moving leftward arrives at $t$. This is represented by the 
first term in Eq. (7). Another wave starts from $t_0$, moving leftward arrives at 
the turning point $t_1$ to get reflected to arrive back at $t$. This is a 
reflected wave contribution given by
\be
\exp{\left[+iS(t_1,t_0)-iS(t,t_1)\right]}=
\exp{\left[2iS(t,t_0)-iS(t,t_0)\right]}.
\ee
The contribution (10) is then multiplied by repeated reflections between the complex 
turning points $t_1$ and $t_2$ and its contribution is
\be
\sum_{\mu=0}^{\infty}\left[ -i\exp{\{iS(t_1,t_2)\}}\right]^{2\mu}=
{1\over {1+\exp{\left[ 2iS(t_1,t_2)\right]}}}.
\ee
The contribution (10) times (11) give the second term in Eq. (7) as the reflected wave 
where $R$ is the reflection amplitude. The essence of Eq. (7) is that there is no 
particle at $t\rightarrow -\infty$ i.e.,
\be
\psi_{in}{\matrix{\longrightarrow \cr t\rightarrow -\infty}}
\exp{\left[iS(t,t_0)\right]},\qquad t\,<\,0
\ee
but at $t\rightarrow +\infty$, Eq. (12) evolves into
\be
\psi_{in}{\matrix{\longrightarrow \cr t\rightarrow +\infty}}
{1\over {(\omega (t))^{1/2}}}
\left[ \exp{[iS(t,t_0)-iR[-iS(t,t_0)]]}\right].
\ee
We identify $R$ as pair production amplitude. The above method works also in 
space dependent gauge in which we are to replace Eq. (12) and Eq. (13) as 
\be
\psi_{in}(x){\matrix{\longrightarrow\cr x\rightarrow\infty}}e^{-iS(x,x_0)}.
\ee
\be
\psi_{in}(x){\matrix{\longrightarrow\cr x\rightarrow -\infty}}e^{-iS(x,x_0)}
-iRe^{+iS(x,x_0)}.
\ee
We applied this technique for particle production by external electromagnetic 
background for $A_\mu=(E_0x,0,0,0)$ as follows. The Klein-Gordon 
equation is
\be
((\partial_\mu+iqA_\mu)(\partial^\mu+iqA^\mu)+m^2)\Phi=0.
\ee
With $M=k_{x}^{2}+k_{y}^{2}+m^2=k_{\bot}^{2}+m^2$ and
\be
\Phi=e^{-i\omega t}e^{ik_x x+ik_y y+ik_zz}\phi(x),
\ee
we find from Eq. (16)
\be
{{\partial^2 \phi}\over {\partial x^2}}+\left[ (\omega+qE_0)^2-M^2\right]\phi=0.
\ee
Let us substitute
\be
\rho=\sqrt{qE_0}x+{\omega\over {q\sqrt{E_0}}},
\ee
\be
\lambda={{k_{x}^{2}+k_{y}^{2}+m^2}\over {qE_0}},
\ee
in Eq. (18) to get
\be
{{\partial^2\phi}\over {\partial x^2}}+(\rho^2-\lambda)\phi=0,
\ee
so that the turning points are at $\rho=\pm\sqrt{\lambda}$. We evaluate $R$ 
using Eq. (9) and Eq. (20) as
\be
\vert R\vert^2={{e^{-\pi\lambda}}\over {(1+e^{-\pi\lambda})^2}}.
\ee
We evaluated \cite{bis:ijtp} the same problem in time dependent gauge, found complex turning 
points in complex time plane to obtain the same expression of $R$. Thus the 
reflection amplitude $R$ in CWKB is a gauge independent result. 
\par
Recently \cite{srini:iucaa,srini1:iucaa} a complex path analysis similar to us 
[they used the technique 
of Landau and needs to know the mode solutions] obtained the result
\be
\vert R_{\sc{C}}\vert^2={{e^{-\pi\lambda}}\over {1+e^{-\pi\lambda}}}.
\ee
This result is also obtained in \cite{brout:pr} but differs from our expression (22). 
Henceforth we call our reflection amplitude as $R_{\sc{CWKB}}$ to distinguish it 
from Eq. (23). In \cite{brout:pr}, the unitarity relation (that also follows from 
charge conservation)
\be
\vert R_{\sc{C}}\vert^2+\vert T_{\sc{C}}\vert^2=1,
\ee
is equated with
\be
\vert\alpha_\lambda\vert^2-\vert\beta_\lambda\vert^2=1
\ee
(here $\alpha_\lambda$ and $\beta_\lambda$ are Bogolubov coefficients) to obtain
\be
\vert\beta_\lambda\vert^2=
{{\vert R_{\sc{C}}\vert^2}\over {\vert T_{\sc{C}}\vert^2}}
\ee
\be
\vert\alpha_\lambda\vert^2={1\over {\vert T_{\sc{C}}\vert^2}}.
\ee
No arguments are placed to obtain Eq. (26) and Eq. (27) though 
$\vert\beta_\lambda\vert^2={{\vert T_{\sc{C}}\vert^2}\over 
{\vert R_{\sc{C}}\vert^2}}$ and
$\vert\alpha_\lambda\vert^2={1\over {\vert R_{\sc{C}}\vert^2}}$ is also a 
possible choice.
It is not at all clear what transpires to connect $\vert R_{\sc{C}}\vert^2$ and 
$\vert T_{\sc{C}}\vert^2$ with $\vert\alpha_\lambda\vert^2$ and 
$\vert\beta_\lambda\vert^2$, wherein one can in principle determine the 
number of pairs formed $(\equiv\vert\beta_\lambda\vert^2)$ simply from the 
results of $\vert R\vert^2$ only. The reason for such a choice may be that one 
knows a priori 
the result of Schwinger to settle Eqs. (26) and (27). In a recent work 
\cite{bis:grg} we call into question the validity of Eqs. (26) and (27). In the 
present work we show that both $\vert R\vert $ and $\vert R_{\sc{C}}\vert $ 
given in Eqs. (22) and (23) are correct, in which $R (\equiv $ full S-matrix 
element) is related to disconnected propagator and $R_{\sc{C}} (\equiv $ 
connected S-matrix element) is related to connected propagator. We also derive the 
relations (26) and (27) from our result of $R$.
\section{\bf{Reflection Coefficient And Average Number Of Pairs}}
The amplitude for observing a scalar particle at the point $x_a$ and an 
antiparticle at $x_b$ is given by \cite{har:prd,bar:ictp}
\be
A=-A_0\int d{\vec{k_a}} d{\vec{k_b}} d\sigma_{a}^{\mu} d\sigma_{b}^{\nu}
f^*(k_a,x_a) \stackrel{\longleftrightarrow}{\partial_{\mu a}}G(x_b,x_a)
\stackrel{\longleftrightarrow}{\partial_{\nu b}}f^*(k_b,x_b)
\ee
where $d\sigma^\mu$ is the element of a space-like hypersurface, 
$\partial_\mu$ is the derivative in the time like direction and $A_0$ is the 
amplitude for no particle production. Consider scalar particle in time dependent 
gauge. In such cases, the $\vec{x}$ dependence of $f$ and $G$ are all phase 
factors and hence the integrations over $d^3x_{a,b}$ (required in evaluating 
the Green function
$G(x_b,x_a)$) and $d^3k_{a,b}$ fix the above numbers of the particle and 
antiparticle to be $-\vec{k_a}=\vec{k_b}=\vec{k}$. If we look at $t_a=t_b$, we 
see a particle and an antiparticle on the same space-like surface with 
opposite momenta. Under these circumstances Eq. (28) can be written as
\be
A=\int d^3kA(\vec{k})
\ee
where $A(\vec{k})$ being the amplitude for creation of a pair with momenta 
$\vec{k} \,and\,-\vec{k}$. For an electromagnetic potential $A_\mu=(0,E_0t,0,0),$ the 
Klein-Gordon equation is
\be
m^2f=(-\partial^{2}_{t}+(\partial_x-qE_0t)^2+\partial^{2}_{y}+\partial^{2}_{z})f.
\ee
With
\be
f(\vec{x},t)={1\over {(2\pi)^{3/2}}}e^{i\vec{k}.\vec{x}} g(t),
\ee
we find
\be
g=ND_{-{1\over 2}-i{\lambda\over 2}}(\sqrt{2iqE_0} (t-{{k_x}\over {qE_0}}),
\ee
where
\be
\vert N\vert=(2qE_0)^{-{1\over 4}}e^{-{{\pi\lambda}\over 8}},
\ee
and
\be
\lambda\equiv {1\over {qE_0}}(m^2+k_{y}^{2}+k_{z}^{2}).
\ee
Evaluating the propagator it can be shown that $A(\vec{k})$ 
turns out to the form
\be
\vert A(\vec{k})\vert^2=\vert A_0\vert^2\omega_k
\ee
with
\be
\omega_k={1\over {1+e^{\pi\lambda}}}.
\ee
Thus $\vert A(\vec{k})\vert$ is the probability for the creation of one pair 
with wave number $\vec{k}$, $\vert A_0\vert^2$ is the absolute probability of 
creating no particles in a given mode (i.e., in mode $k$) and $\omega_k$ is the 
relative probability of producing a pair if none were present in the initial 
state. This $\omega_k$ is the square of the connected S-matrix element 
\cite{cor:prd} and is denoted by $R_{\sc{C}}$.
\par
It should be pointed out that in the complex path analysis, we could start from 
the relation (35) itself to determine $\vert A_0\vert^2$ and $\omega_k$ using 
the result of $R_{\sc{CWKB}}$ with 
$\vert A(\vec{k})\vert^2\equiv \vert R_{\sc{CWKB}}\vert^2$. Henceforth we omit 
the subscript `CWKB' i.e., write $R_{\sc{CWKB}}\equiv R$. It may be pointed 
out that $R$ is  the full $S$-matrix elements whereas the $\omega_k$ by 
definition is related only to the
connected $S$-matrix elements so that $\omega_k=\vert R_C(\vec{k})\vert^2$ i.e.,
$R_C$ is related to the connected propagator $S_{AC}$ whereas $R$ is related to 
the disconnected propagator $S_A$. These propagators are related by
\be
S_A=e^{iW}S_{\sc{AC}}
\ee
where $W=Tr\,\ln{S_{\sc{AC}}}$ is the sum of connected vacuum graphs 
[see \cite{cor:prd}].
\par
The probability of producing $n$ pairs of particles with wave number $\vec{k}$ is 
then 
\[ {P_n(\vec{k})=\vert A_0\vert^2\omega_{\vec{k}}^{n}.}\]
Hence the total probability of creating $n=0,1,2,\ldots$ pair will sum to unity 
(for Bose statistics) i.e., 
\beq
\vert A_0\vert^2(1+\omega_{\vec{k}}+\omega_{\vec{k}}^{2}+\ldots )= 1 \nonumber \\
or, \;\vert A_0\vert^2 = 1-\omega_{\vec{k}}.
\eeq
Hence the absolute probability for 1-pair production is
\be
\vert A(\vec{k})\vert^2\equiv P_1(\vec{k})=(1-\omega_{\vec{k}})\omega_{\vec{k}}.
\ee
By definition the reflection coefficients 
$\vert R_{\sc{CWKB}}\vert^2\equiv \vert R\vert^2$ equals Eq. (39) so that
\be
\vert R\vert^2=\omega_k(1-\omega_k).
\ee
The average number of pairs with wavenumber $\vec{k}$ is
\be
N({\vec{k}})=\sum_{n=0}^{\infty}nP_n({\vec{k}})={{\omega_{\vec{k}}}\over {1-\omega_{\vec{k}}}}.
\ee
Let us now relate Eqs. (38), (40) and (41) to the vacuum persistence probability. By 
definition $\vert A_0\vert^2$ is the probability of no pair production for wave 
number $\vec{k}$. Hence the total probability of no pair production for all 
wave number $\vec{k}$ is
\beq
\prod_{\vec{k}}\vert A_0(\vec{k})\vert^2&=&\prod_{k}(1-\omega_k)
=\prod_{k}(1+N(\vec{k}))^{-1} \nonumber \\
&\equiv & e^{-\sum_{\vec{k}}\ln{(1+N(\vec{k}))}}.
\eeq
Eq. (42) gives the probability that no particles are created in the vacuum in 
any given mode i.e., the probability that the vacuum remains the vacuum is given 
by
\beq
\vert <out\vert in >\vert^2&=&\prod_{\vec{k}}\vert A_0(\vec{k})\vert^2 \nonumber \\
or,\, e^{-2Im\,{{\cal L}_{eff}}. V_4}&=&e^{-\sum_{k}\ln{(1+N(\vec{k}))}}\nonumber\\
or,\,2Im\,{{\cal L}_{eff}}V_4&=&\sum_{\vec{k}}\ln{(1+N(\vec{k}))},
\eeq
where $V_4=L^3T.$
In CWKB
\be
\vert R\vert^2={{e^{-\pi\lambda}}\over {(1+e^{-\pi\lambda})^2}}.
\ee
Using Eq. (40), we find two roots
\be
\omega_{\vec{k}}={1\over {1+e^{-\pi\lambda}}},\;\;
{{e^{-\pi\lambda}}\over {1+e^{-\pi\lambda}}}.
\ee
We take the second root so that
\be
\omega_{\vec{k}}={{e^{-\pi\lambda}}\over {1+e^{-\pi\lambda}}}.
\ee
\be
1-\omega_{\vec{k}}={1\over {1+e^{-\pi\lambda}}}.
\ee
The reasons for such a choice will shortly be clear. Using the 
values of Eqs. (46) and (47) in Eq. (41), we get
\be
N({\vec{k}})=e^{-\pi\lambda}
\ee
so that
\be
2Im{{\cal L}_{eff}}.V_4=\sum_{k}\ln{(1+e^{-\pi\lambda})}.
\ee
If the electric field $E_0\rightarrow 0$, $\vert out>\rightarrow\vert in>$
and there is no instability so that $<out\vert in>=1$. This implies that 
$Im{{\cal L}_{eff}}\rightarrow 0$. From (34), we see
$\lambda={{(m^2+k_{\bot}^{2})}\over {qE_0}}\rightarrow \infty$ when 
$E_0\rightarrow 0$. Hence from Eq. (40) we have $Im{{\cal L}_{eff}}=0$ implying 
$<out\vert in>=1$. Had we chosen the other root in Eq. (45), we would get
$N({\vec{k}})=\exp{(\pi\lambda)}$ and it goes to infinity as $E_0\rightarrow 0$,
giving $<out\vert in>\rightarrow infinite$. This is unphysical. Thus the 
correct choice is given by Eq. (46).
\par
Expanding the log term in Eq. (43) we get 
\beq
V_4.2Im{{\cal L}_{eff}}&=&
\sum_{{\vec{k}},n}(-1)^{n+1}{1\over n}
e^{-\pi n\lambda}\nonumber \\
&=&
\sum_{{\vec{k}},n}(-1)^{n+1}{1\over n}
e^{{-\pi n(k_{\bot}^{2}+m^2)}\over {qE_0}}\nonumber .
\eeq
We convert the sum by integral as follows
\beq
2Im{{\cal L}_{eff}}V_4
&=&\sum_{n=1}^{\infty}
\int dk_y {L\over{2\pi}}
\int dk_z {L\over{2\pi}}
\int_{0}^{qE_0L} dk_0 {T\over{2\pi}}
e^{{-\pi n(k_{\bot}^{2}+m^2)}\over {qE_0}}(-1)^{n+1}{1\over n}\nonumber \\
&=&{{(qE_0)^2L^3T}\over {(2\pi)^3}}\sum_{n=1}^{\infty}(-1)^{n+1}{1\over {n^2}}
e^{{{-n\pi m^2}\over {qE_0}}}.
\eeq
Since $V_4=L^3T$, we get
\be
2Im{{\cal L}_{eff}}
={{(qE_0)^2}\over {(2\pi)^3}}\sum_{n=1}^{\infty}(-1)^{n+1}{1\over {n^2}}
e^{{{-n\pi m^2}\over {qE_0}}}.
\ee
This is exactly the Schwinger result for $Im{{\cal L}_{eff}}$ in one loop 
approximation. Eq. (43) is the Nikishov's representation (Eq. (2)). Our derivation no 
where requires to evaluate the Bogolubov co-efficient or to know the mode 
solutions. This amply reflects the gauge invariant content of CWKB.
\section{\bf{Reflection Coefficient And Bogolubov Coefficients}}
Let the positive and negative frequency solutions be denoted by $\phi_{k\pm}$
(in the limit $t\rightarrow -\infty$) and 
\be
\Phi=\sum_{{\vec{k}}}( a_{{\vec{k}}}^{in} \phi_{{\vec{k}}(+)}
+ b_{{\vec{k}}}^{in} \phi_{{\vec{k}}(-)}),
\ee
\be
a_{{\vec{k}}}^{in}\vert Vac>\equiv
a_{{\vec{k}}}^{in}\vert in>=0.
\ee
Eq. (53) specifies the in vacuum. For out vacuum let us define
\be
\Phi=\sum_{{\vec{k}}}(
a_{{\vec{k}}}^{out}
\phi_{{\vec{k}}}^{(+)}
+ b_{{\vec{k}}}^{out}
\phi_{{\vec{k}}}^{(-)}),
\ee
\be
a_{{\vec{k}}}^{out}\vert out>=0
\ee
where $\phi_{{\vec{k}}}^{(\pm)}$ are the positive and negative frequency 
solutions at $t\rightarrow +\infty$. The Bogolubov coefficients 
$\alpha_{{\vec{k}}}$
$\beta_{{\vec{k}}}$
are now given by 
\be
\Phi_{{\vec{k}}(+)}= 
\alpha_{{\vec{k}}} \phi_{{\vec{k}}}^{(+)}+
\beta_{{\vec{k}}} \phi_{{\vec{k}}}^{(-)},
\ee
\be
\Phi_{{\vec{k}}(-)}= 
\beta_{{\vec{k}}}^{*} 
\phi_{{\vec{k}}}^{(+)}+
\alpha_{{\vec{k}}}^{*} \phi_{{\vec{k}}}^{(-)},
\ee
with
\[{\vert\alpha_{{\vec{k}}}\vert^2- \vert\beta_{{\vec{k}}}\vert^2=1,}\]
\[{\phi_{{\vec{k}}}^{(+)}=
[\phi_{{\vec{k}}}^{(-)}]^*,\;
\phi_{{\vec{k}}(+)}=
[\phi_{{\vec{k}}(-)}]^*.}\]
The operators $ a_{{\vec{k}}}^{in}$ and $b_{{\vec{k}}}^{in}$ are connected to the
out operators as (we now omit the vector sign over $k$)
\be
a_{k}^{in}=
\alpha_{k}^{*}a_{k}^{out}-\beta_{k}^{*}b_{k}^{out},
\ee
\be
b_{k}^{+in}=
-\beta_ka_{k}^{out}+\alpha_kb_{k}^{+out}.
\ee
The pair formation amplitude is 
\be
a_{pair}=<out\vert a_{k'\,out}b_{{\vec{k}}\,out}\vert in\,>.
\ee
Using (59), we write it as
\beq
a_{pair}
&=&
<out\vert a_{k'\,out}
(\alpha_{k}^{*-1} b_{{\vec{k}}\,in}
+{{\beta_{k}^{*}}\over {\alpha_{k}^{*}}}
a_{k\,out}^{+})\vert in>
\nonumber \\
&=&
{{\beta_{k}^{*}}\over {\alpha_{k}^{*}}}
<out\vert a_{k'\,out}
a_{k\,out}^{+}\vert in\,>\nonumber \\
&=&
{{\beta_{k}^{*}}\over {\alpha_{k}^{*}}}
<out\vert \delta_{kk'}+a_{k\,out}^{+}
a_{k'\,out}\vert in\,>=
{{\beta_{k}^{*}}\over {\alpha_{k}^{*}}}<out\vert in>,
\eeq
since $[a_{k\,out},a_{k'\,out}^{+}]=\delta_{kk'}$ and 
$<out\vert a_{k}^{+}=0$. The relative amplitude of producing a pair if 
none were present in the initial state is
\be
{{<\,out\vert a_{k'\,out}b_{k\,out}\vert in\,>}\over {<\,out\vert in\,>}}.
\ee
The square of this amplitude gives the probability of creating a pair in the 
given mode. It is $\omega_k$ as denoted earlier [see Eq. (36) and the 
discussion thereafter]. Hence, using (61) and (62) 
\be
\omega_k=
\vert {{\beta_k}\over {\alpha_k}}\vert^2
=\vert R_{\sc{C}}\vert^2.
\ee
We thus get
\be
{{\omega_k}\over {1-\omega_k}}=
{{\vert{\beta_k}\vert^2}\over {\vert\alpha_k\vert^2-\vert\beta_k\vert^2}}
=\vert\beta_k\vert^2.
\ee
Hence
\be
\vert R_{\sc{C}}\vert^2={{e^{-\pi\lambda}\over {1+e^{-\pi\lambda}}}}.
\ee
This result of $R_{\sc{C}}$ was derived in \cite{srini:iucaa,srini1:iucaa} using 
the technique of Landau \cite{lan:quant} and also in \cite{brout:pr} by saddle 
point method. Let us try to understand the relation between $R$ and $R_{\sc{C}}$.
\beq
\vert R\vert^2&=&{1\over {1+e^{-\pi\lambda}}}.\vert R_{\sc{C}}\vert^2\nonumber\\
&=&e^{-\ln{(1+N(k))}}\vert R_{\sc{C}}\vert^2\nonumber\\
&=&e^{ -2Im\,{{\cal L}_{eff}}(k).V_4}
\vert R_{\sc{C}}\vert^2.
\eeq
This connection between $R$ and $R_{\sc{C}}$ was also cited in ref. 
\cite{cor:prd} 
where
\beq
2Im\,{{\cal L}_{eff}}V_4&=&2\sum_{k}Im\,{{\cal L}_{eff}}(k).V_4\nonumber\\
&=&\sum_{k}\ln{(1+N(k))}.
\eeq
From (63), we find
\be
\vert R_{\sc{C}}\vert^2={{\vert\beta_k\vert^2}\over {\vert\alpha_k\vert^2}},
\ee
so that
\be
\vert T_{\sc{C}} \vert^2 =1-\vert R_{\sc{C}}\vert^2=
{{\vert \alpha_k\vert^2-\vert\beta_k\vert^2}\over
{\vert\alpha_k\vert^2}}={1\over{\vert\alpha_k\vert^2}}.
\ee
From the unitarity relation $\vert R_{\sc{C}}\vert^2+\vert T_{\sc{C}}\vert^2 =1$, 
we get
\be
\vert T_{\sc{C}}\vert^2={1\over {\vert\alpha_k\vert^2}}.
\ee
Hence
\be
\vert\beta_k\vert^2={{\vert R_{\sc{C}}\vert^2}\over {\vert T_{\sc{C}}\vert^2}},\;
\vert\alpha_k\vert^2={1\over {\vert T_{\sc{C}}\vert^2}}
\ee
The results (70) and (71) coincide with (26) and (27) as were taken in 
\cite{brout:pr}. The CWKB result of $R$ thus correctly reproduces the result of 
$R_{\sc{C}}$ obtained by other methods.
\section{\bf{Conclusion}}
Let us try to synopsize the results. In the complex path analysis \cite{lan:quant}, that uses 
Landau's technique or in the saddle point method \cite{brout:pr}, one obtains $R_{\sc{C}}$ 
and $T_{\sc{C}}$. These $R_{\sc{C}}$ and $T_{\sc{C}}$ are related by charge 
conservation as follow                           
\be
\vert R_{\sc{C}}\vert^2 +\vert T_{\sc{C}}\vert^2=1,\qquad (scalar\;\,case)
\ee
\be 
1+\vert R_{\sc{C}}\vert^2=\vert T_{\sc{C}}\vert^2,\qquad (spinor\;\,case).
\ee
Using the unitarity constraint of Bogolubov coefficients, one determines 
$\vert\beta_{k}\vert^2$ with the identification $N(k)=\vert\beta_k\vert^2$. 
Using Nikishov representation one relates $N(k)$ with $Im\,{{\cal L}_{eff}}$ to 
obtain the Schwinger result. Here one needs to know the mode solutions as well 
as, the continuation from $t\rightarrow -\infty \,(x\rightarrow -\infty)$ to 
$t\rightarrow +\infty \,(x\rightarrow +\infty)$. Moreover in 
\cite{brout:pr,lan:quant}, there is an 
ambiguity in fixing $\beta_k$ and $\alpha_k$ from Eqs. (72) and (73). In CWKB, 
we calculate $R$ by simple integrals and particle and antiparticle states are 
fixed by WKB definition (Parker's definition). This fixed the vacuum. Nikishov's 
and Schwinger's results follow through our derivation. The instability of the 
vacuum that causes particle production occurs due to motion in complex-coordinate 
plane. There is a rotation of currents as the system evolves from 
$t\rightarrow -\infty\;(x\rightarrow -\infty)$ to $t\rightarrow +\infty\;
(x\rightarrow +\infty)$. The mixing of positive and negative frequency states 
occurs between the turning points. The value of the wavefunction at real $t$ is 
also contributed, not only by real trajectories but also by complex 
trajectories, in the sense of WKB approximation. Our result shows that CWKB 
justifiably takes at least one loop quantum correction within the semiclassical 
approximation. We have shown elsewhere in detail that the particle production 
in CWKB is equivalent to tunneling from left Rindler wedge to right Rindler 
wedge via real $x$ and imaginary time plane. Not realizing the difference 
between $R$ and $R_{\sc{C}}$ and the beauty of the complex path analysis, some 
circles argue the expression of $R$ as wrong comparing it with $R_{\sc{C}}$. We 
clarify our view in this respect and establish the correctness of CWKB.

\section{\bf{Acknowledgment}}                          
\par
A.Shaw acknowledges the financial support from ICSC World Laboratory,
LAUSSANE during the course of the work. The authors are thankful to Prof P.
Dasgupta for a critical reading of the manuscript and for discussion during the
preparation of this work. A part of the work was done during one of the authors'
(S. Biswas) stay at Inter University Centre for Astronomy and Astrophysics, Pune,
India.
\end{document}